\documentstyle[sprocl,epsfig]{article}

\def\be{\begin{equation}}
\def\ee{\end{equation}}
\def\bea{\begin{eqnarray}}
\def\eea{\end{eqnarray}}

\begin{document}

\title{
MULTIPARTICLE PRODUCTION IN HIGH-MASS DIFFRACTION DISSOCIATION
AND THE INTERPLAY OF PHOTONS AND POMERONS
\footnote{Talk presented by R.~Engel at PHOTON'97, Egmond aan Zee, 
The Netherlands, May 10 - 15, 1997}
}


\author{F.W.~Bopp}
\address{Universit\"at Siegen,
Fachbereich Physik, D--57068 Siegen, Germany}

\author{R.~Engel}
\address{DESY, D--22603 Hamburg, Germany
}

\author{J.~Ranft}
\address{INFN - Laboratori Nationali del Gran Sasso, I--67010 Assergi AQ,
Italy
}

\author{A.~Rostovtsev}
\address{Institute of Theoretical and Experimental Physics
Moscow, Russia}


\maketitle

\vspace*{-8cm}
\begin{flushright}
Siegen University SI-97-10\\
June 1997
\end{flushright}
\vspace*{+7cm}

\abstracts{
Multiple interaction models satisfying $s$-channel unitarity
predict that, in contrast to inelastic processes, factorization is 
violated in diffractive processes.
The size of this effect can be characterized in terms of the
rapidity gap survival probability.
The possibility of its measurement at HERA is pointed out.
Furthermore a method to measure photon diffraction dissociation at LEP2
and planned linear colliders is discussed and cross section 
predictions are given.
}


\section{Unitarity, pomerons, and factorization}

Assuming that high virtual masses are damped due to the dynamics of the
the strong interaction, hadronic interactions can be described
by Gribov's Reggeon field theory (RFT)\cite{Gribov67a-e,Baker76}.
The total amplitude can be written as the sum of $n$-pomeron exchange
amplitudes $A^{(n)}(s,t)$.
Unitarity implies that at high energies graphs with $n$-pomeron exchange
become important. However, it should be emphasized that only the
one-pomeron exchange graph satisfies factorization as assumed, for
example, in parton model calculations of hadronic jet production.

Why do we expect factorization in inclusive processes?
For example, let's consider the simplest ``factorization-breaking''
contribution, the two-pomeron graph. 
\begin{figure}[!htb]
\centering
\hspace*{0.25cm}
\psfig{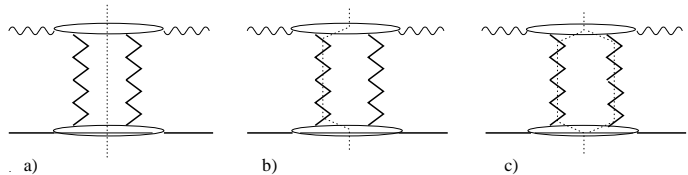}
\caption{
\label{pom2-cut} \em
Breakdown of the total discontinuity of the two-pomeron exchange graph
according to the AGK cutting rules:
a) the diffractive cut describing low-mass diffraction,
b) the one-pomeron cut, and
c) the two-pomeron cut.}
\end{figure}
To discuss particle production, we apply the optical theorem
together with the AGK cutting rules\cite{Abramovski73-e}.
Three different
cut configurations are giving the dominant
contributions:
the diffractive cut with the weight 1
(Fig.~\ref{pom2-cut} a)), the one-pomeron cut with the weight -4
(Fig.~\ref{pom2-cut} b)),
and the two-pomeron cut with the weight 2 (Fig.~\ref{pom2-cut} c)).
\begin{figure}[!htb]
\centering
\hspace*{0.25cm}
\psfig{file=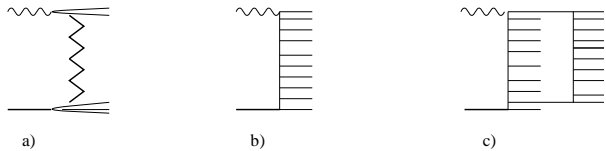,width=80mm}
\caption{
\label{pom2-ine} \em
Inelastic final states resulting from 
a) the diffractive cut describing low-mass diffraction,
b) the one-pomeron cut, and
c) the two-pomeron cut.}
\end{figure}
Assuming that a two-pomeron cut gives two times the particle
yield compared to the one-pomeron cut (in central pseudorapidity
region, see Fig.~\ref{pom2-ine}),
the inclusive inelastic charged particle cross section reads
\begin{equation}
\frac{d\sigma_{\rm ch}}{d\eta}\bigg|_{\eta\approx 0}
\sim
 0\times (+1) \frac{dN_1}{d\eta}
+ 1\times (-4) \frac{dN_1}{d\eta}
+ 2\times (+2) \frac{dN_1}{d\eta}
= 0
\label{AGK-cancellation}
\end{equation}
where the particle density in pseudorapidity of produced by a 
one-pomeron cut is denoted by $dN_1/d\eta$.
Note that the cross section contribution of the two-pomeron graph
vanishes. Analogously, the factorization violating contributions 
due to multi-pomeron exchange graphs cancel out in all orders. 
This means that only the one-pomeron graph
determines the inclusive particle cross section in the central region
(AGK cancellation).
It can be shown that the same cancellation effects hold true also 
in the case of inclusive jet production.

In high-mass diffraction dissociation we have to consider only final state
configurations with sufficiently large rapidity gaps. 
It is important to notice that all the configurations with
more than one cut pomeron (multiple-interaction contributions, see
Fig.~\ref{pom2-ine} c)) are not
considered for the diffractive cross section since in this case the
rapidity gap of the diffractive process is filled by
particles belonging to additional pomeron cuts.
However, as shown above,
these configurations are needed to cancel other negative terms implied
by unitarity.
Consequently, factorization is violated in diffraction dissociation
since the cross section contributions of the higher-order 
multi-pomeron graphs do not vanish.
For example, within the triple-pomeron approximation the diffractive cross 
section would grow with the 
energy like $\sigma_{\rm diff} \sim s^{2 \Delta}$. 
The measured
flat energy dependence is explained due to unitarity corrections: 
additional interactions produce particles filling the rapidity gap of the 
diffractive interaction. This can be effectively parametrized introducing
a energy- and process-dependent
{\it rapidity gap survival probability} $\langle|S|^2\rangle$
\cite{Bjorken93a}.

\section{A possible measurement of $\langle|S|^2\rangle$}

In the following we will discuss some consequences for
particle production in high-mass photon dissociation at HERA.
In comparison to hadron-hadron interactions, there are two 
important new effects to note: {\bf (i)} the photon has a dual
nature and can interact as a gauge boson (pointlike) or a hadron
(resolved), and {\bf (ii)} the photon has an additional degree of freedom, the
photon virtuality.
\begin{figure}[!hbt]
\begin{center}
\unitlength1mm
\begin{picture}(115,42)
\put(0,-3){\psfig{file=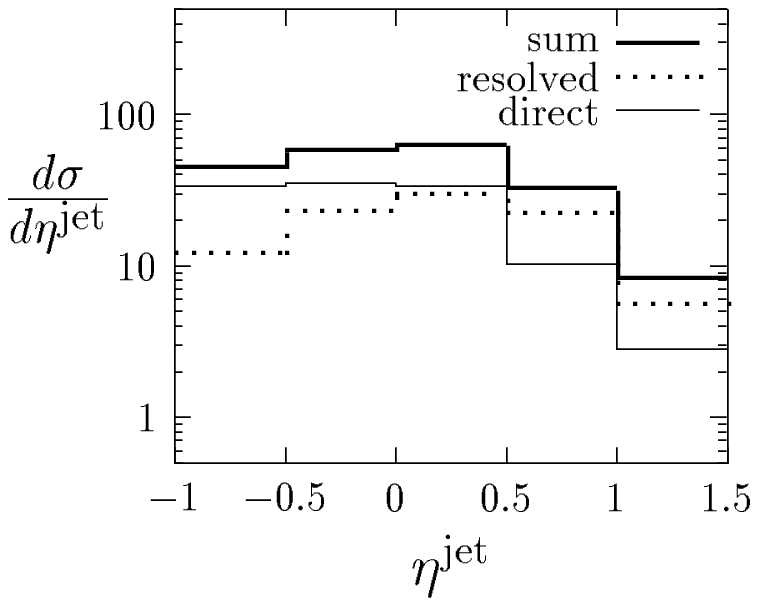,width=5.5cm}}
\put(60,-3){\psfig{file=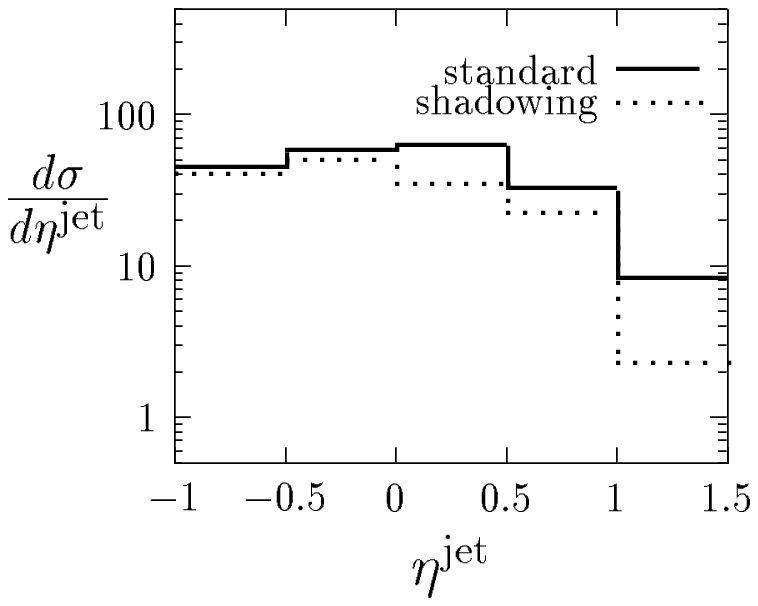,width=5.5cm}}
\end{picture}
\end{center}
\caption{\label{diff-jet4}
Differential single-inclusive jet cross section $E_\perp > 6$ GeV: a) 
breakdown in direct and resolved processes for $\langle |S_{\rm res}|^2
\rangle$ =1,
b) cross sections for $\langle |S_{\rm res}|^2 \rangle$ =1 (standard) and
$\langle |S_{\rm res}|^2 \rangle$ =0.5 (shadowing)
}
\end{figure}
Both (i) and (ii) give a handle to suppress
the relative size of the unitarity corrections (e.g.\ the relative size
of the multi-pomeron graphs compared to the one-pomeron exchange)
\cite{Bopp96a}.

Using these effects, the rapidity gap survival
probability can be determined experimentally as follows\\
{\bf (i)} measurement of the diffractive structure function $F_3^D$ in
deep-inelastic photon-proton scattering\\
{\bf (ii)} extraction of parton densities of the pomeron (gluon densities are
determined from scaling violation effects {\it without} using
photoproduction data)\\
{\bf (iii)} application of these parton densities to the calculation of
single-inclusive particle cross sections or jet cross sections in
high-mass diffraction dissociation of real photons, comparison with
measurements to determine $\langle |S_{\rm res}|^2 \rangle$.\\
In $\gamma^\star p$ scattering with
not too small $Q^2$, multiple pomeron
exchange contributions are suppressed at least by a factor $1/Q^2$
compared to the leading amplitude. Hence in diffractive DIS
unitarity corrections are small and the measurement (i) provides the
``true'' parton density of the pomeron.
However, in photoproduction unitarity effects (e.g.\ multiple-pomeron
exchange contributions) are important. Furthermore the rapidity
gap survival probability in hard diffraction differs significantly 
between direct and resolved photon interactions. 
In direct photon interactions, there is no hadronic remnant to allow for
multiple interactions (e.g. two-pomeron exchange).
Rapidity gap events with a resolved hard photon interaction are suppressed 
by a factor of about $2\dots 3$ compared to events with direct hard photon
interactions. 
Having the possibility to distinguish in experiment between diffractive 
direct and
resolved photoproduction in (iii) 
offers a unique means to check the
predictions of multiple-interaction models and the concept of the
rapidity gap survival probability. For example, we consider
single-inclusive jet production applying the cut $\eta_{\rm max} < 1.5$
to select diffractive events in a simple ``black--white'' model where 
all resolved processes are treated as purely hadronic ones. 
The calculations were done using {\sc Phojet} \cite{Engel95a,Engel95d}.
In Fig.~\ref{diff-jet4} a) the cross
section is shown for direct and resolved events separately assuming
$\langle |S_{\rm res}|^2 \rangle$ =1 for all processes. 
In Fig~\ref{diff-jet4} b)
the total jet  cross section is shown for the case of $\langle |S_{\rm
res}|^2 \rangle$ =1 and $\langle |S_{\rm res}|^2 \rangle$ =0.5. It
should be emphasized that both curves differ in shape. Similar results
are obtained in case of the transverse energy distribution of jets where
resolved processes contribute mainly to the low-$E_\perp$ part.

\section{On the determination of the diffractive cross section in
$\gamma\gamma$ collisions}

The diffractive contribution to the total
$\gamma\gamma$ cross section is difficult to measure since the LEP
detectors have only a very small acceptance for such events. On the
other hand, the knowledge of the diffractive cross section is very
important for many theoretical calculations as well as background
estimations.

\begin{figure}[!hbt]
\begin{center}
\unitlength1mm
\begin{picture}(115,40)
\put(0,-3){\psfig{file=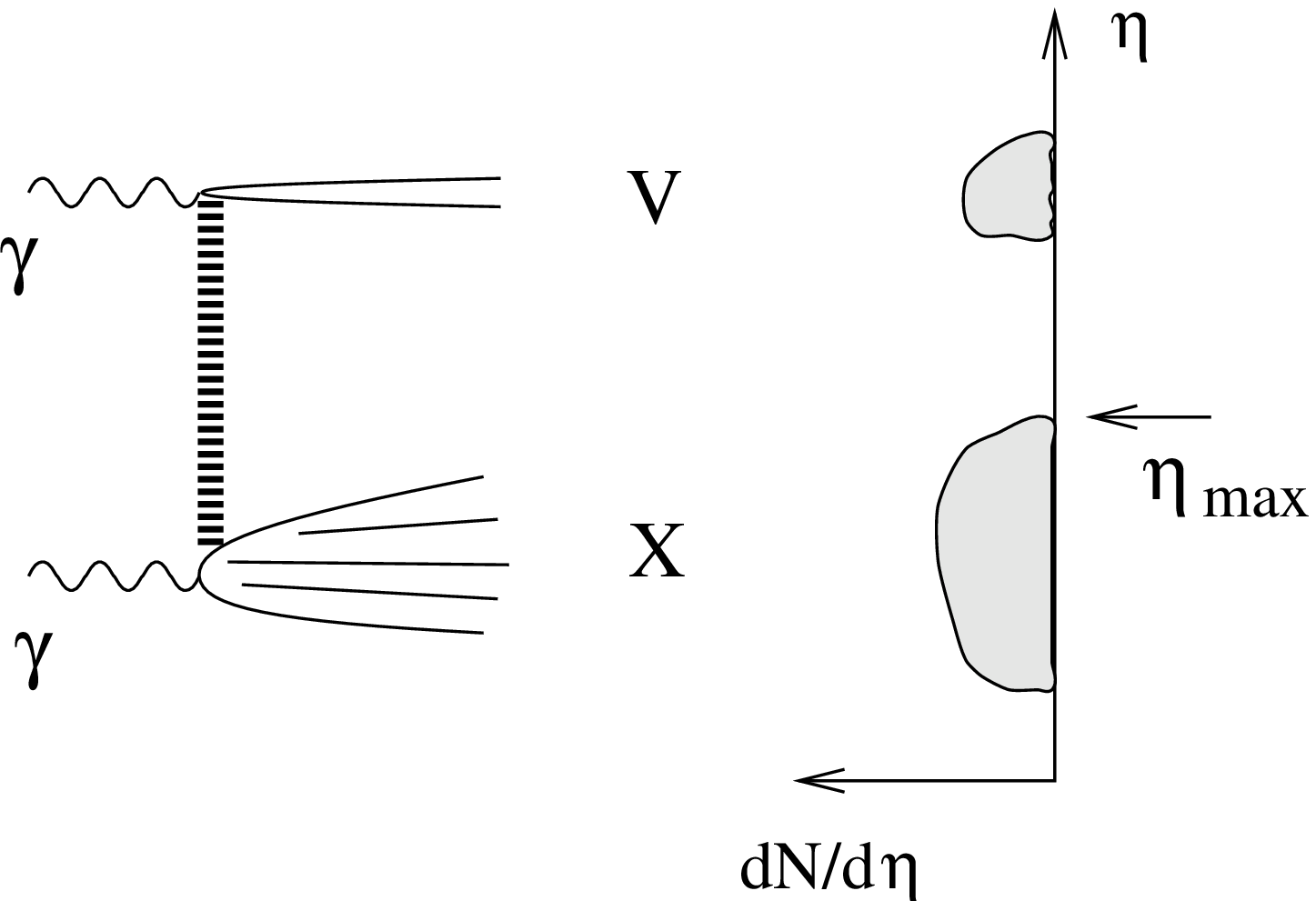,width=5.5cm}}
\put(60,-3){\psfig{file=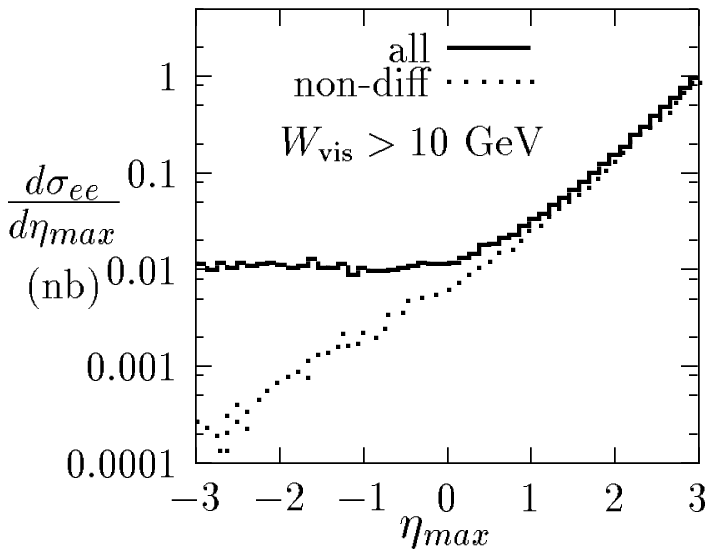,width=5.5cm}}
\end{picture}
\end{center}
\caption{\label{etamax-LEP}
a) definition of the variable $\eta_{\rm max}$ in $\gamma\gamma$
collisions.
b) $ee$ cross section for all and for non-diffractive $\gamma\gamma$ at
LEP2.
}
\end{figure}

In analogy to the $\eta_{\rm max}$ cut applied by the HERA
collaborations to identify diffractive events, a similar quantity can be
defined for the case of $\gamma\gamma$ interactions \cite{Engel96j}. 
Here one has to
deal with the variation of the rapidity of the $\gamma\gamma$ system
in the lab. frame. A possible way to define $\eta_{\rm max}$ in this
case is to use a forward em. detector as a trigger for hadronic
activity. In events with forward hadronic activity, $\eta_{\rm max}$ is
then given by the pseudorapidity of the most-forward scattered particle
seen in the central detector (for example, with a coverage of $|\eta| <
3$). This is illustrated in Fig.~\ref{etamax-LEP} a).
The differential $\eta_{\rm max}$ cross section is shown in
Fig.~\ref{etamax-LEP} b). Selecting events with a visible energy
$W_{\rm vis}>10$ GeV leads to an almost flat cross section for negative
values of $\eta_{\rm max}$ clearly showing the diffractive contribution
to the total $\gamma\gamma$ cross section. Of course, the method
presented here can also be applied to $\gamma\gamma$ collisions at linear
colliders.
\\[2mm]
We acknowledge valuable discussions with S.~Roesler.
One of the authors (R.E.) was supported in parts by the Deutsche
Forschungsgemeinschaft under contract No.\ Schi 422/1-2.


\section*{References}


\begin{thebibliography}{1}

\bibitem{Gribov67a-e}
V.~N. Gribov:
\newblock Sov.\ Phys.\ JETP 26 (1968) 414

\bibitem{Baker76}
M.~Baker and K.~A. Ter-Martirosyan:
\newblock Phys.\ Rep.\ 28C (1976) 1

\bibitem{Abramovski73-e}
V.~A. Abramovski, V.~N. Gribov  and O.~V. Kancheli:
\newblock Sov.\ J.\ Nucl.\ Phys.\ 18 (1974) 308

\bibitem{Bjorken93a}
J.~D. Bjorken:
\newblock Phys.\ Rev.\ D47 (1993) 101

\bibitem{Bopp96a}
F.~W. Bopp, R.~Engel  and A.~Rostovtsev:
\newblock Hadron production in photon collisions at high energies,
\newblock (hep-ph/9612344) to appear in Proceedings of the XXVI International
  Symposium on Multiparticle Dynamics, Faro, Portugal,  1996

\bibitem{Engel95a}
R.~Engel:
\newblock Z.\ Phys.\ C66 (1995) 203

\bibitem{Engel95d}
R.~Engel and J.~Ranft:
\newblock Phys.\ Rev.\ D54 (1996) 4244

\bibitem{Engel96j}
R.~Engel and A.~Rostovtsev:
\newblock How to measure diffraction in two-photon collisions at LEP,
\newblock Univ.\ Siegen preprint SI 96-12, (hep-ph/9611205),  1996

\end{thebibliography}


\end{document}